\documentclass[preprint,12pt]{elsarticle}

\usepackage{amssymb}
\usepackage{amsmath}

\usepackage{listings}
\usepackage{xcolor}
\usepackage{tikz}
\usetikzlibrary{calc}
\usepackage{upquote} 

\newcounter{bla}

\journal{Computer Physics Communications}

\begin{document}

\begin{frontmatter}



\title{Symmetr: a Python package for determining symmetry properties of crystals}


\author[a]{Jakub Železný\corref{author}}

\cortext[author] {Corresponding author.\\\textit{E-mail address:} zeleznyj@fzu.cz}
\address[a]{Institute of Physics of the Czech Academy of Sciences, Cukrovarnická 10/112 162 00 Praha 6 }

\begin{abstract}
Condensed matter compounds typically form crystals, which break the rotational and translational invariance of space but remain invariant under a discrete set of symmetry operations. Understanding the effects allowed by this symmetry breaking, as well as the constraints imposed by the crystal structure, is a crucial problem in condensed matter physics. Here, we present a Python package for determining the symmetry-restricted forms of tensors describing physical properties of crystals, focusing particularly on magnetic materials. The primary focus is on response tensors; however, the program can also describe equilibrium properties and other physical properties, such as magnetic interactions. The program can describe the symmetry using the conventional magnetic space groups, as well as using the spin groups that describe the non-relativistic limit. Additional functionality includes the treatment of quantities projected onto a particular site and expansions in the magnetic order parameter. The code can be used either from the command line or via a Python API.
\\


\noindent \textbf{PROGRAM SUMMARY}

\begin{small}
\noindent
{\em Program Title: Symmetr}                                          \\
{\em CPC Library link to program files:} (to be added by Technical Editor) \\
{\em Developer's repository link:} https://github.com/zeleznyj/symmetr \\
{\em Licensing provisions(please choose one):} MPL-2.0  \\
{\em Programming language: Python}                                   \\
{\em Supplementary material:}                                 \\
{\em Nature of problem:}\\
  Symmetr is a Python package designed to determine symmetry-restricted forms of tensors describing physical properties of crystalline materials. Based on Neumann’s principle, it automates the otherwise complex task of enforcing crystal and magnetic symmetries on equilibrium and response tensors, which is essential for analyzing transport, magnetic, and other material properties. The package supports both non-magnetic and magnetic systems, including cases with broken time-reversal symmetry, and extends naturally to the non-relativistic limit through the use of spin groups, enabling symmetry analysis in the absence of spin–orbit coupling.\\
{\em Solution method:}\\
  The program first determines the symmetry operations of the crystal and then the transformation of the tensor under the symmetry operations. This then leads to a series of linear equations. We solve these using the singular value decomposition followed by gaussian ellimation. The program is implemented in Python and is used from the command line or through a Python API.
\end{small}
   \end{abstract}
\end{frontmatter}

\section{Introduction}
Symmetry is a crucial concept in condensed matter physics. Crystals break the homogeneity and isotropy of space, as well as, in some cases, spatial inversion or time-reversal symmetry. This symmetry breaking results in a wide variety of phenomena. For example, breaking inversion symmetry leads to the spin-orbit torque \cite{Manchon2019}, a torque acting on the magnetic order induced by an electrical current. The breaking of time-reversal symmetry in magnetic materials leads to the anomalous Hall effect \cite{RevModPhys.82.1539,Smejkal2022}. However, crystals typically remain invariant under a discrete set of symmetry operations, which restrict their response properties. For instance, whereas all ferromagnets exhibit the anomalous Hall effect, only some antiferromagnets do \cite{Smejkal2022}, depending on the symmetry of the crystal. 

The symmetry properties of crystals can be determined based on Neumann's principle, which states that if a crystal is invariant under certain symmetry elements, any of its physical properties must also be invariant under the same symmetry elements. This principle constrains the form of both equilibrium and response properties of crystals. These properties are typically represented by tensors, and determining the symmetry-restricted form of such tensors is an important topic in condensed matter physics. While this can be done manually in simple cases, it becomes impractical or impossible for more complex problems, making software tools necessary.

Many common cases have been tabulated. The symmetry groups of all non-magnetic crystals fall within the 230 crystallographic space groups. However, for physical phenomena describing global properties of crystals, translational symmetry is irrelevant. In such cases, it is sufficient to consider only the 32 crystallographic point groups. Furthermore, for physical phenomena that are invariant under spatial inversion, it suffices to consider only the 11 Laue groups. This significantly simplifies symmetry classification and has enabled tabulation for many phenomena.

The focus of this work is mainly on magnetic materials, for which the situation is more complex. The classification of symmetries in magnetic materials requires the inclusion of time-reversal symmetry, which increases the number of symmetry groups. There are 1651 magnetic space groups, 122 magnetic point groups, and 32 magnetic Laue groups. While tabulating the symmetry of physical phenomena for magnetic materials is still possible and has been done for many properties \cite{Kleiner1966, Wimmer2016, Gallego:lk5043}, but it becomes less practical. 

Magnetic space groups implicitly assume the presence of spin-orbit coupling, a relativistic interaction that couples the spin degree of freedom to the lattice. Although spin-orbit coupling exists in all materials, it is often weak and can be treated perturbatively or even neglected. It is therefore useful to consider symmetry in the absence of spin-orbit coupling, corresponding to the non-relativistic limit. In this case, symmetry is described by the so-called spin groups \cite{Brinkman1966,Litvin1974}. Recently, a classification of spin space groups has been published \cite{Xiao2024}; however, the number of such groups is very large, making exhaustive tabulation of symmetry-restricted physical properties impractical.

In condensed matter research, one is typically interested in determining the symmetry of a particular physical property for a given crystal with a specific magnetic structure. This is the problem addressed by the Symmetr package. Full classification is not required; instead, it is sufficient to determine the relevant symmetry operations and symmetrize the given tensor accordingly. This approach also enables the treatment of local phenomena that depend on site-group symmetry or on expansions in the magnetic order parameter, for which knowledge of the space group alone is insufficient. This is necessary, for example, for describing spin-orbit torques in antiferromagnetic systems. The approach offers additional advantages; for instance, starting from a given crystal structure allows the coordinate system to be fixed consistently with respect to the crystal.

\section{Methodology}

To determine the symmetry-restricted form of a tensor describing some physical phenomenon for a given crystal, three steps are necessary:

\begin{enumerate}
    \item Determine the symmetry operations.
    \item Determine the transformation of the tensor by the symmetry operations.
    \item Find the most general tensor shape that satisfies the restrictions imposed by the symmetry operations.
\end{enumerate}

For determining the symmetry operations in the case where spin-orbit coupling is included (i.e., magnetic space groups), the Symmetr package uses the Findsym program \cite{findsym,findsym_website} internally. In the non-relativistic case, a custom algorithm is implemented, which determines all the symmetry operations. This is described in detail in Section \ref{section:noso_symmetry}.

To determine how a given tensor transforms under a symmetry operation, it is necessary to consider its constituent relations. Within magnetic space groups, there are 4 different symmetry operation types to take into account: rotation, spatial inversion, time-reversal, and translation. When using spin groups, we have to additionally consider pure spin rotations.

\subsection{Transformation under rotation}

We consider as an example the conductivity tensor $\sigma$, which is defined by:

\begin{align}
\mathbf{j} = \sigma \mathbf{E},\label{eq:conductivity_tensor}
\end{align}
where $\mathbf{j}$ is the current and $\mathbf{E}$ is the applied electric field.
Because the laws of physics are invariant under rotations, it must hold for a rotation $R$:

\begin{align}
R \mathbf{j} = \sigma^R R \mathbf{E},
\end{align}
where $\sigma^R$ is the conductivity tensor of the rotated system. This equation can be intuitively understood as rotating the whole experiment, which cannot change the result. Multiplying this equation by $R^{-1}$ and writing the indices explicitely we have:

\begin{align}
j_i = R^{-1}_{ik}R^T_{jl}\sigma^R_{kl} E_j.
\end{align}
This equation must hold for all $\mathbf{j}$ and $\mathbf{E}$. By comparing with Eq. \eqref{eq:conductivity_tensor} we thus find

\begin{align}
\sigma_{ij} = R^{-1}_{ik}R^T_{jl}\sigma^R_{kl} \iff
\sigma^R_{ij} = R_{ik}R^{-T}_{jl} \sigma_{kl}.
\end{align}
Here $R^{-T}$ denotes the inversion and transpose of the matrix $R$. This thus gives the transformation of the conductivity tensor. We see that the two indices of the tensor transform differently. This is referred to as covariant and contravariant indices of the tensor: the index $j$ is contravariant, and the index $i$ is covariant.  The difference comes from how they transform under a change of the coordinate system. The difference between covariant and contravariant indices only matters in non-Cartesian coordinate systems, however. In a Cartesian (orthonormal) coordinate system, they transform the same. This can be seen from the fact that in a Cartesian coordinate system $R^{-T} = R$.

A general tensor corresponding to a global property of the crystal will transform the same:

\begin{align}
\chi^R_{i\ldots j,k\ldots l} = R_{im}\ldots R_{jn} R^{-T}_{ko} R^{-T}_{lp} \chi_{m\ldots n, o \ldots p},
\end{align}
where the indices $i\ldots j$ are contravariant and the indices $k\ldots l$ are covariant.

\subsection{Transformation under inversion}

The same approach can also be applied to the spatial inversion. Since the inversion is represented by the matrix
\begin{align}
\mathbf{P} = \left(\begin{matrix}
-1 & 0 & 0 \\
0 & -1 & 0 \\
0 & 0 & -1
\end{matrix}\right)
\end{align}
there is no difference between covariant and contravariant indices. However, the indices will transform differently depending on whether they correspond to a polar or an axial vector: polar vectors change sign under inversion, whereas axial vectors do not. For example, taking the conductivity tensor, since both $\mathbf{j}$ and $\mathbf{E}$ are polar vectors, both indices of the tensor change sign under inversion, which means that the tensor as a whole is invariant under inversion.

\subsection{Transformation under translations \& atomic permutations}

For tensors corresponding to global phenomena, the translations do not matter. However, when we are interested in local properties, such as a torque on a specific magnetic site, translations have to be taken into account as well. The important information is how a particular symmetry transforms the individual atoms, which we refer to as atomic permutation. Considering a symmetry operation $R$ that transforms atom $a$ into atom $b$, it holds for a site-projected response tensor $\chi_a$

\begin{align}
 \chi^R_b = \chi_a
\end{align}
where $\chi^R$ denotes the transformation of the tensor by the symmetry operation. Since the translations do not matter directly, they are not stored within the code; only the permutation of the atomic positions is.

\subsection{Time-reversal symmetry}

Time-reversal is a symmetry that reverses the direction of time in the microscopic equation of motion. Consequently, it also reverses both orbital and spin angular momentum. This is why it is used as additional symmetry for the classification of magnetically ordered crystals. The tensor transformation for the time-reversal symmetry cannot be derived from the constituent relations, like for the crystallographic symmetry operations. This issue has been noticed already by Birs \cite{BUTZAL1982518}. Taking the conductivity tensor, for example, the electric field is even under the time-reversal, whereas the electrical current is odd. Since non-magnetic crystals are invariant under time-reversal symmetry, we would then find that there is no conductivity allowed by symmetry in non-magnetic crystals, which is clearly wrong. The issue with this naive approach has been discussed by Shtrikman and Tomas \cite{Shtrikman1965} and later by Grimmer \cite{Grimmer1993}. The core of the issue lies in the fact that currents are dissipative processes for which reversing the microscopic paths of motion makes no physical sense.

The theory of dissipative processes has been studied by Onsager, who has derived the so-called Onsager relations \cite{PhysRev.37.405}. For conductivity, these state

\begin{align}
 \sigma_{ij}([\mathbf{M}],\mathbf{H}) = \sigma_{ji}(-[\mathbf{M}],-\mathbf{H})
\end{align}
where $[\mathbf{M}]$ denotes the magnetic configuration (i.e., the magnetic moments or magnetization density) and $\mathbf{H}$ the external magnetic field. This thus means that the symmetric component of the conductivity tensor is time-reversal even, whereas the anti-symmetric is odd. This gives us a prescription on how to transform the tensor by the time-reversal symmetry: we first split the tensor into the symmetric and anti-symmetric components and then transform each by the time-reversal separately.

The conductivity is a special case in that the Onsager relations relate the tensor to itself. In general cases, the Onsager relations relate the effect to its inverse. For example, the electric current induced by a thermal gradient and a thermal current due to a voltage. In such a case, the Onsager relations do not restrict the form of the response tensor itself. However, we can still separate any tensor into the time-reversal even and time-reversal odd components

\begin{align}
\chi^e([\mathbf{M}]) &= \frac{\chi(\mathbf{[M]}) + \chi(\mathbf{-[M]})}{2},\\
\chi^o([\mathbf{M}]) &= \frac{\chi(\mathbf{[M]}) - \chi(\mathbf{-[M]})}{2}.
\end{align}
This is useful in practice since these two components typically behave differently and are often separated both in theory and in experiment. For example, for the spin-orbit torque, the time-reversal odd component corresponds to the field-like torque, whereas the time-reversal even component is the anti-damping-like torque. 
The transformation under time-reversal can also be derived from the microscopic formulas. The time-reversal operator in quantum mechanics

\begin{align}
 \hat{\mathcal{T}} = iK\sigma_y,
\end{align}
where $K$ denotes complex conjugation and $\sigma_y$ is the $y$ Pauli matrix. By directly transforming the Kubo formula for the linear response, it is possible to see that it contains two components that have an opposite transformation since one component has a real part of a matrix element, whereas the other has an imaginary part. This approach has been used in several works \cite{Kleiner1966,Wimmer2016,Zelezny2017}.

\subsection{Spin rotations}

In the magnetic space groups, all real space rotations are coupled with the corresponding spin space rotation. Since mirrors correspond to a combined rotation and inversion, and spin is invariant under inversion, mirrors are coupled with the spin rotation corresponding to the rotation part of the mirror. The indices corresponding to spin will transform by the spin-rotation matrix. In the case of the spin-groups, the real and spin space parts are decoupled, and thus the real space transformation can be combined with any spin-rotation.

\subsection{Magnetic interactions}

Magnetic interactions are commonly described with a Heisenberg-like Hamiltonian of the form:

\begin{align}
    H = \sum_{ab,ij} H^{(2)}_{ab,ij} M^a_i M^b_j +
    \sum_{abcd,ijkl} H^{(4)}_{abcd,ijkl} M^a_i M^a_j M^a_k M^a_l + \dots,
\end{align}
where $M^a_i$ is the $i$-the component of the magnetic moment on site $a$. The $H^{(n)}_{ab\dots,ij\dots}$ tensors are expansion coefficients corresponding to the various magnetic interactions whose symmetry we need to determine. The symmetrization of these tensors follows the same principle as the response tensors; however, since it has some peculiarities, we discuss it here separately.

To determine the symmetry of the $H^{(n)}_{ab\dots,ij\dots}$ tensors, we have to consider the symmetry operations of the non-magnetic crystals. This is because these are symmetry operations that relate different magnetic configurations of the magnetic crystal. Considering a symmetry operation $\hat{R}$ that transforms magnetic moments such that the moment on site $a$ is given by $R_{ij} M^{P^{-1}(a)}_j$ where $R_{ij}$ is the matrix representing transformation of magnetic moment under $\hat{R}$ (that is transformation of a time-reversal odd axial vector) and $p$ is the permutation of atoms under $R$. This means that under $R$ the atom $a$ transforms to atom $p(a)$ and atom $p^{(-1)}(a)$ to $a$.  All the magnetic configurations related by some symmetry of the non-magnetic lattice must have the same energy. Considering the second-order for simplicity, it must thus hold

\begin{align}
    \sum_{ab,ij} H^{(2)}_{ab,ij} M^a_i M^b_j = \sum_{ab,ij} H^{(2)}_{ab,ij} R_{ik} M^{p^{-1}(a)}_k R_{jl} M^{p^{-1}(b)}_l 
\end{align}
We can replace the same over $a$ and $b$ by a sum over $p(a)$ and $p(b)$ since both run over all atomic sites. Then we get

\begin{align}
    \sum_{ab,ij} H^{(2)}_{ab,ij}  = \sum_{ab,ij}R_{ik}^{T}R_{jl}^{T} H^{(2)}_{P(a)P(b),kl}  M^{a}_i  M^{b}_j, 
\end{align}
where we have also transposed the $R$ matrices and interchanged the summing indices $i \leftrightarrow k$ and $j \leftrightarrow l$. Since this must hold for all $M^{a}$ we get

\begin{align}
    H^{(2)}_{ab,ij} M^a_i M^b_j = R_{ik}^{T}R_{jl}^{T} H^{(2)}_{P(a)P(b),kl}.
\end{align}
This thus gives a transformation rule for the tensor $H^{(2)}$. Higher orders can be derived completely analogously. Additionally, we have to apply a constraint that the tensor has to be symmetric under interchanging the indices corresponding to the same atom. 

For the symmetrization of the $H^{(n)}_{ab\dots,ij\dots}$ tensors, it is necessary to keep track of atomic permutations. Here, there is an important limitation with the applicability of the code. Within the code, we do not keep track of the unit cell to which the atom belongs. That is, we do not distinguish between a transformation that transforms atom $a$ to atom $b$ in the same unit cell and a transformation that transforms atom $a$ to atom $b$ in a different unit cell. This means that we cannot distinguish between a symmetry of, for example,  $H^{(2)}_{ab,ij}$ corresponding to $a$ and $b$ in the same unit cell and $a$ and $b$ in different unit cells. Therefore, the code can describe the magnetic interactions only in the so-called macrospin approximation, in which the sample is composed only of a single domain, and the same atom in different unit cells all have the same magnetic moment directions and can thus be treated as equivalent. It is possible to specify two atoms on the same positions as non-equivalent, which turns off the symmetrization for these indices. This is sufficient in most cases to treat the interactions for atoms in different cells; however, a full general treatment of interactions in different unit cells is not supported.

\subsection{Expansion in magnetization}

For magnetic systems, the response tensors can be expanded in the magnetic order parameter. Within the code, this is implemented only for collinear systems, which are described by a single order parameter $\mathbf{n}$:

\begin{align}
\chi_{ijk}(\mathbf{n}) = \chi_{ijk}^{(0)} + \chi^{(1)}_{ijkl} n_l + \chi^{(2)}_{ijklm} n_l n_m + \dots
\end{align}
The transformation of the expansion tensors can be determined similarly to that of the magnetic interactions. Similarly to the magnetic interactions case, we have to consider the non-magnetic symmetry operations. In single-sublattice ferromagnets, the order parameter is the net magnetization, which transforms as a time-reversal odd axial vector. In systems with more magnetic sublattices, the situation can be more complex, and the transformation of the order parameter has to be determined based on the permutation of the atoms.

\subsection{Non-relativistic symmetry}
\label{section:noso_symmetry}

In the absence of spin-orbit coupling, the symmetry is described by the so-called spin-groups \cite{Brinkman1966,Litvin1974}. These differ from the magnetic space groups that describe the relativistic symmetry in that the spin and spatial rotations are decoupled. That is, with spin-orbit coupling, a rotation is composed of a real-space rotation and an equivalent spin rotation. Without spin-orbit coupling, however, the spin rotation can be different from the real-space part.

In a non-magnetic system, the spin group is a product of a crystallographic space group, the group containing all spin rotations and time-reversal. In a magnetic system, the symmetry is reduced. For example, magnetic systems are not invariant under time-reversal or an arbitrary spin-rotation.

Since all the symmetry operations of the magnetic system must also be symmetry operations of the non-magnetic system, we use the symmetry operations of the non-magnetic system as a starting point. We assume that the symmetry of the system is described by assigning magnetic moments to each site. This is usually satisfied; however, it may also happen that the whole magnetization density needs to be taken into account. In such a case, the symmetry determined based on the approach used here would be higher than in the real system. However, the general principles outlined here would still apply.

The procedure for obtaining the non-relativistic symmetry operations is as follows:

\begin{enumerate}
 \item Identify all the non-magnetic symmetry operations. For this, we use the Findsym code. We consider only the spatial part of the symmetry operations, and we need to find any spin rotations that, together with the spatial component, are symmetries of the magnetic system.
 \item Because the spatial and spin transformations are decoupled, the only aspect of the spatial symmetry operations that matters for the spin rotation is how they permute the magnetic atoms. That is, under a combined spin-rotation $R_s$ and spatial transformation $R$, the magnetic moments of the system will transform as $\mathbf{M}_a \rightarrow R_s \mathbf{M}_{p(a)}$, where $p(a)$ denotes the permutation corresponding to $R$. The combined operation will be a symmetry of the system if for all moments $\mathbf{M}_a \rightarrow R_s \mathbf{M}_{p(a)}$.

 For example, a symmetry operation may transform atom $A$ to atom $B$, atom $B$ to atom $C$, and atom $C$ to atom $A$. We denote this operation as $ A \rightarrow B \rightarrow C \rightarrow A$ and refer to it as a permutation chain. Each permutation chain must end with the same atom as the one it starts with. Each symmetry operation is described by one or more such chains.

 \item Now for every permutation chain we have to find the spin-rotations that leave it invariant, that is, spin-rotations such that $R_s \mathbf{M}_{c_i} = \mathbf{M}_{c_{i+1}}$, where $\mathbf{M}_{c_i}$ denotes the magnetic moment of each chain. This has to be done separately with and without time-reversal. In the following, we assume that all the moments have the same magnitude, since otherwise they cannot be symmetry related.

 Since the last element of the chain must be the same as the first, we have a condition:
 \begin{align}
 &\textrm{Without time-reversal:}\quad \mathbf{M}_{A} = R_s^n \mathbf{M}_{A}\\
 &\textrm{With time-reversal:}\quad \mathbf{M}_{A} = (-1)^n R_s^n \mathbf{M}_{A}
\end{align}
where $n$ is the number of unique atoms in the chain (that is, for chain $ A \rightarrow B \rightarrow C \rightarrow A$, the length is 3). From this, it is possible to determine the necessary conditions for the existence of a spin-rotation that leaves the chain invariant.

If the chain is collinear, then the symmetry analysis is quite simple. The only two relevant symmetries are an arbitrary rotation around the collinear axis $R_s^{||}$ and a $\pi$ rotation around any perpendicular axis $R_s^\perp$. The system can either be invariant under $R_s^{||}$ or under $R_s^\perp R_s^{||}$. In this case, the chain has a continuous spin-rotation symmetry.

For a non-collinear system, the conditions for the rotation angle $\theta$ are:

\begin{itemize}
 \item Without time-reversal or with time-reversal and $n$ even:  $\theta = \frac{i \pi}{n}$, where $i$ is an even integer.
 \item With time-reversal and $n$ odd: $\theta = \frac{i\pi}{n}$, where $i$ an odd integer.
 \end{itemize}

\item For each permutation pair in the chain, we now have to determine the spin-rotations that leave it invariant, taking into account the restriction on $\theta$. We do this by first considering each pair separately. That is, for example, for the chain $ A \rightarrow B \rightarrow C \rightarrow A$ we have to find spin-rotations $R_s$ that satisfy $R_s A = B$, $R_s B = C$, and $R_s C = A$ and then find the spin-rotations that are common to all pairs.

In general, it can be shown that a rotation $R_s$ with an angle $\theta$ connecting two vectors $\mathbf{M}_A$ and $\mathbf{M}_B$, will exist if $\mathbf{M}_A \cdot \mathbf{M}_B \geq \cos(\theta)$, that is the angle between the two vectors must greater or equal to $\theta$. The rotation axes will be given by:

\begin{align}
 \mathbf{n}^{1+} &= \cos(\alpha) \frac{\mathbf{M}_A + \mathbf{M}_B}{||\mathbf{M}_A + \mathbf{M}_B||}  + \sin(\alpha)  \frac{\mathbf{M}_A \times \mathbf{M}_B}{||\mathbf{M}_A \times \mathbf{M}_B||},\notag\\
 \mathbf{n}^{1-} &= -\cos(\alpha) \frac{\mathbf{M}_A + \mathbf{M}_B}{||\mathbf{M}_A + \mathbf{M}_B||}  - \sin(\alpha)  \frac{\mathbf{M}_A \times \mathbf{M}_B}{||\mathbf{M}_A \times \mathbf{M}_B||}.\notag \\
 \mathbf{n}^{2+} &= \cos(\alpha) \frac{\mathbf{M}_A + \mathbf{M}_B}{||\mathbf{M}_A + \mathbf{M}_B||}  - \sin(\alpha)  \frac{\mathbf{M}_A \times \mathbf{M}_B}{||\mathbf{M}_A \times \mathbf{M}_B||},\label{eq:n_final}\\
 \mathbf{n}^{2-} &= -\cos(\alpha) \frac{\mathbf{M}_A + \mathbf{M}_B}{||\mathbf{M}_A + \mathbf{M}_B||}  + \sin(\alpha)  \frac{\mathbf{M}_A \times \mathbf{M}_B}{||\mathbf{M}_A \times \mathbf{M}_B||},\notag
\end{align}

where
\begin{align}
 \alpha = \arccos\left( \sqrt{\frac{2 \left( \mathbf{M}_A \cdot \mathbf{M}_B - \cos(\theta) \right)}{(1-\cos(\theta))(1 + \mathbf{M}_A \cdot \mathbf{M}_B)}} \right).
\end{align}

The $\pm$ axes correspond to an opposite sense of rotation. In general, only one of the two is correct. It can easily be checked which one is the correct one for a given $\mathbf{M}_A$, $\mathbf{M}_B$, and $\theta$.

\item Once we have the spin-rotations that are symmetries of each permutation chain, we find the spin-rotations (if any) that are common to all chains. These, together with the spatial part of the symmetry corresponding to the permutation and potentially time-reversal, are symmetries of the system.

\end{enumerate}

\subsection{Symmetrizing tensors}

Once we have the symmetries for a given system and the transformation of the tensor by the symmetries, we can determine the general symmetry-restricted shape of the given tensor by solving a set of linear equations imposed by the symmetries. For a given symmetry $R$, it must hold that

\begin{align}
 \chi^R_{ijk\dots} = \chi_{ijk\dots}
 \label{eq:chi_R_eq_chi}
\end{align}
We denote the independent components of the tensor using variables $x_{ijk\dots}$. At the start of the symmetrization procedure, $\chi_{ijk\dots} = x_{ijk\dots}$. In general, the components of both $\chi$ and $\chi^R$ are linear combinations of $x_{ijk\dots}$. Eq. \ref{eq:chi_R_eq_chi} can thus be rewritten as

\begin{align}
 Y\mathbf{x} = 0,
 \label{eq:Yx}
\end{align}
where $\mathbf{x}$ is a vector of all $x_{ijk\dots}$ variables and Y is a numerical matrix with dimension $3^n \times 3^n$ where $n$ is the rank of the tensor $\chi$. This is a system of linear equations, which in general either has only a null solution or has infinitely many solutions. In the former case, $x_{ijk\dots} = 0$ and thus the tensor is null. In the latter case, if the number of solutions is lower than the number of independent variables, we can eliminate some of the independent variables. This can be conveniently achieved by using Gaussian elimination to transform the matrix into the reduced row echelon form; however, in practice, this method is numerically unstable and is thus not suitable. Instead, we use the SVD decomposition to determine the basis of the null space of $Y$, that is, the basis of solutions of Eq. \eqref{eq:Yx}. We use these basis vectors to construct a new matrix $\tilde{Y}$

\begin{align}
 \tilde{Y} = \left(
    \begin{matrix}
        \mathbf{a}_1 \\
        \mathbf{a}_2 \\
        \vdots
    \end{matrix}
  \right),
\end{align}
where $\mathbf{a}_1$, $\mathbf{a}_2$, \dots are the basis vectors of the null space. For matrix $\tilde{Y}$ it must also hold that $\tilde{Y}\mathbf{x} = 0$. Since the vectors $\mathbf{a}_1$, $\mathbf{a}_2$, \dots are linearly independent, we can now safely use the Gaussian elimination to eliminate the dependent variables.

\section{Using the code}

Here, a brief description of how the code is used is given. The full documentation of the code's features is distributed with the code and available at \cite{symmetr_docs}. When using the code in scientific publications, cite the paper as well as Findsym \cite{findsym,findsym_website}.

To install the code, use pip:

\begin{verbatim}
    pip install symmetr
\end{verbatim}
or alternatively, download the code from GitHub and run 

\begin{verbatim}
    pip install .
\end{verbatim}
from the directory where the code is located. There are tests distributed with the code. To run the tests, use pytest:
\begin{verbatim}
cd tests
pytest
\end{verbatim}

The code can be used from the Linux command line or alternatively can be used from Python. The code has two modes. The first mode is the response mode that allows determining the symmetry of response or equilibrium tensors defined by the observables of the phenomena of interest. The other mode is for determining the symmetry of Heisenberg-like magnetic Hamiltonians.

\subsection{Response mode}

To use the response mode, the user has to specify the observable corresponding to the perturbing field and the response observable. The user also has to specify either the crystal structure or the symmetry group, though the functionality of the latter case is rather limited. The observables can be, for example, 'j' for charge current, 'E' for electric field, or 's' for spin. The observables can also be combined, so, for example, 's.v' is a spin current, and when 'E.E' is used for the perturbation, it denotes a second-order response to the electric field. The program always outputs the time-reversal even and odd parts separately. Note that the order of the two can change. The order is always such that the second tensor is the one that corresponds to the intrinsic transformation of the observables under time-reversal, whereas the first has an additional minus sign for symmetries with time-reversal.

Example of using the response mode:

\begin{verbatim}
 symmetr res j E -f Fe.in
\end{verbatim}

Here, 'res' specifies the response mode, 'j' specifies that the response observable is charge current, 'E' specifies that the perturbation is the electric field, and the '-f' switch specifies the input file. Example of the input file for bcc Fe is:
\begin{lstlisting}
Fe
0.001  (*@  \textit{<-- tolerance} @*)
1
3 0 0
0 3 0 (*@  \textit{<-- lattice vectors} @*)
0 0 3
2
I (*@  \textit{<-- centering} @*)
1 (*@  \textit{<-- number of atoms} @*)
Fe (*@  \textit{<-- Atom types} @*)
magnetic
0 0 0 0 0 1 (*@  \textit{<-- Positions and magnetic moments} @*)
\end{lstlisting}
This file serves as an input to Findsym, which is used internally within the program to determine the magnetic symmetry group. The full description of the input file is given in the documentation.

The output of the program for this case is

\begin{verbatim}
even part of the response tensor:
[x00   0    0 ]
[             ]
[ 0   x00   0 ]
[             ]
[ 0    0   x22]
odd part of the response tensor:
[ 0   -x10  0]
[            ]
[x10   0    0]
[            ]
[ 0    0    0]
\end{verbatim}
Here, the first tensor corresponds to normal conductivity, whereas the second is the anomalous Hall effect.

\subsubsection{Atomic projections}

It is possible to evaluate a symmetry of quantities projected to a particular atom, using the '-p' flag. For example

\begin{verbatim}
symmetr res s E -f CuMnAs.in -p 5
\end{verbatim}
gives projected spin-polarization due to the electric field (typically referred to as the inverse spin-galvanic effect or the Edelstein effect) on site 5. The file CuMnAs.in can be found in the Supplementary Material. This flag works by using only symmetries that keep the selected atom invariant. It is also possible to find the symmetry relation (if there is any) between two sites by using the flag '-p2':

\begin{verbatim}
symmetr res s E -f CuMnAs.in -p 5 -p2 6
\end{verbatim}

\subsubsection{Expansion in the order parameter}

It is also possible to evaluate a tensor corresponding to an expansion of a response tensor in the order parameter. This is only implemented for collinear systems, where the order parameter is the spin axis. To enable this mode, use the flag \lstinline|--exp n|, where $n$ is the order of the expansion. For example:

\begin{verbatim}
symmetr res j E -f Fe.in --exp 1
\end{verbatim}
specifies the first-order expansion of the conductivity tensor, which corresponds to the lowest-order expansion term of the anomalous Hall effect. This gives the following result
\begin{verbatim}
[   0      -m2*x210  m1*x210 ]
[                            ]
[m2*x210      0      -m0*x210]
[                            ]
[-m1*x210  m0*x210      0    ]
\end{verbatim}
Here [$m0$,$m1$,$m2$] is the order parameter, which is simply the magnetization in this case.

\subsection{Magnetic Hamiltonian mode}

To use the magnetic Hamiltonian mode, it is necessary to specify the keyword 'mham', the input file, and the sites for which the Hamiltonian is evaluated using the '--sites' keyword. For example

\begin{verbatim}
 symmetr --mham -f Fe.in --sites 1,1
\end{verbatim}
gives

\begin{verbatim}
Hamiltonian term in matrix form:
[x00   0    0 ]
[             ]
[ 0   x00   0 ]
[             ]
[ 0    0   x00]

    2           2           2    
M_1x *x00 + M_1y *x00 + M_1z *x00
\end{verbatim}
The matrix is only printed for the second order and corresponds directly to the $H^2_{11}$ matrix. The second part is the expression $H^2_{11,ij}M^1_iM^1_jM$, which is printed for all orders. This interaction corresponds to uniaxial anisotropy when the moments are in the same cell and to the exchange when they are in different cells. To specify that two atoms on same position belong to different unit cells, it is possible to specify one atom number with a prime, for example,

\begin{verbatim}
 symmetr --mham -f Fe.in --sites 1,1'
\end{verbatim}

The only effect of this flag is that the indices corresponding to these two sites are not enforced to be symmetrical when interchanged.

\subsection{Python API}

All the functionality of the code can be used directly from Python, which allows further processing or automation. The main functions to use are \lstinline|input.parse()|, which parses the input from a command-like input into an internal format, and functions \lstinline|funcs_main.sym_res| and \lstinline|funcs_main.sym_mham|, which are used for the response and magnetic Hamiltonian mode, respectively. For example, to get the conductivity of Fe:

\begin{lstlisting}[language=Python]
from symmetr import sym_res, parse

res = sym_res(parse('res j E -f Fe.in'))
\end{lstlisting}
The \lstinline|res| is a list containing the even and odd response components, which are each represented by a 'Tensors' class. This uses the Sympy library for symbolic mathematics to represent the symbolic tensors. The tensor can be printed using the \lstinline|.pprint()| method:

\begin{lstlisting}[language=Python]
res[0].pprint()
\end{lstlisting}
The \lstinline|pprint()| function has an optional parameter \lstinline|latex|, which turns on printing in LaTeX format.  The individual components can be accessed with normal Python indexing. They are SymPy \cite{sympy} symbolic expressions, and the functionality of SymPy can thus be utilized. Further documentation on the API is available in the documentation of the code \cite{symmetr_docs}.

\subsection{Coordinate system}

The code uses, by default, a Cartesian coordinate system. When using Findsym input, this coordinate system is defined with respect to the lattice as defined in the Findsym input. In case the lattice vectors are explicitly given in the input, the coordinate system is simply the Cartesian coordinate system with respect to which the lattice vectors are defined. In case the type 2 lattice input is used, in which only the lattice vector lengths and angles are given, the Cartesian coordinate system is chosen such that the $\hat{z}$ axis is oriented along the $c$ lattice vector and the $\hat{y}$ axis lies in the $b-c$ plane. It is possible to specify a different coordinate system using the \lstinline|-b| parameter.

\section{Code implementation}

\begin{figure}
    \includegraphics[width=0.95\textwidth]{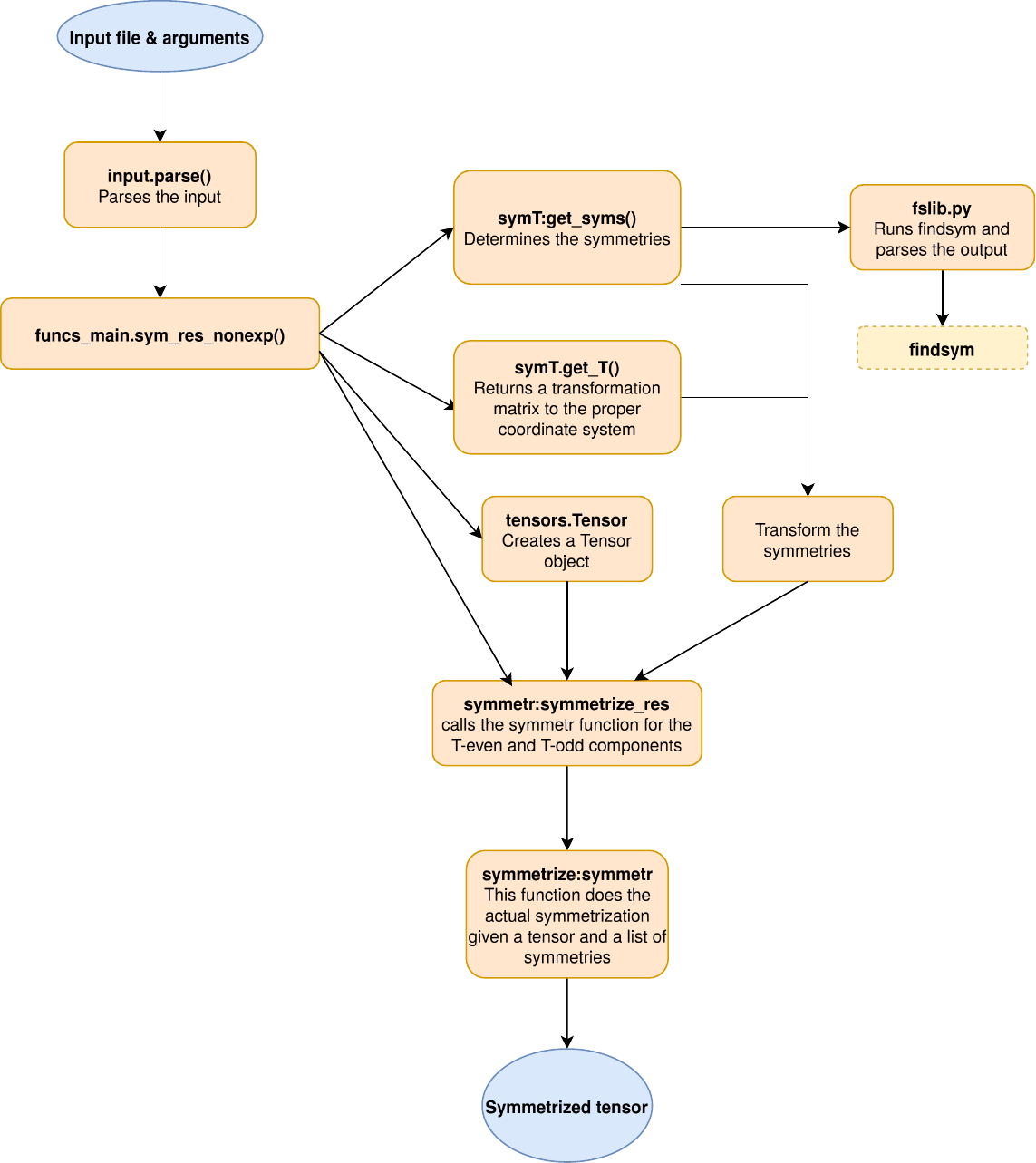}
    \label{fig:flowchart}
    \caption{Simplified flow of the program for response mode without magnetization expansion.}
\end{figure}

The code is written in Python. To determine the symmetries for a given crystal structure and magnetic order, the Findsym package is used. We use the SymPy \cite{sympy} library for symbolic mathematics, which allows us to represent tensors as symbolic objects. An illustration of a typical program flow for the case of a response tensor without magnetization expansion is given in Fig. \ref{fig:flowchart}. The main function that orchestrates the entire symmetrization is \lstinline|funcs_main.sym_res_nonexp(inp)|. This takes as an input the \lstinline|input.options| object created by the \lstinline|input.parse()| function. Analogous functions for response with expansion and for the magnetic Hamiltonian mode are \lstinline|funcs_main.sym_res_exp(inp)| and \lstinline|funcs_main.sym_mham(inp)| respectively. These work very similarly.

The symmetries are obtained using the function \lstinline|symT.get_syms(inp)|. This calls findsym if the input file is used, or determines symmetries based on a group number. Analogous function \lstinline|symT.get_syms_noso(inp)| returns the symmetries for the non-relativistic case. The symmetries are stored using the class \lstinline|symmetry.Symmetry|. The symmetries are returned in the conventional coordinate system, which is defined based on the magnetic symmetry group. Note that even in the non-relativistic case, the coordinate system of the magnetic space group, corresponding to the relativistic symmetry, is used. This coordinate system is dependent on the magnetic order and can be non-orthogonal, which is usually not what the user wants. We thus transform the symmetry operations to a Cartesian coordinate system, which has a fixed orientation with respect to the lattice. It is possible to specify a different coordinate system, although that's rarely needed. Rather than transforming the symmetries, it is also possible to keep the symmetries in the conventional coordinate system and transform the symmetrized tensor. The tensors in both approaches are equivalent, but typically expressed differently. Since transforming the symmetries generally results in tensors expressed more simply and naturally, this is used by default.

Based on the user input, the tensor to be symmetrized is defined. The tensors are represented using the \lstinline|tensors.Tensor| class. This represents a symbolic tensor. This class is practical for output formatting and for further manipulation of tensors, and thus it is used for output. However, for numerical efficiency, a class \lstinline|tensors.NumTensor| is implemented and used for the symmetrization. This is used for representing tensors in which each component is a linear combination of the symbolic variables. Then the tensor can be represented as a numerical matrix that specifies the coefficient of each symbolic variable and each tensor component. For the tensor, it is necessary to specify the transformation rules, which is done using the method \lstinline|tensors.Tensor.def_trans()|.

The symmetrization is done using the \lstinline|symmetrize.symmetr()| function. This function takes as input a tensor, a list of symmetries, and a function that transforms the tensor by the symmetries. In the case of the response tensors, this function is \lstinline|tensors.transform()|. The \lstinline|symmetrize.symmetr()| function is called from the function \lstinline|symmetrize.symmetrize_res()|, which defines the transformation function, selects symmetries or the projection based on the user input. The \lstinline|symmetrize.symmetr()| function is called twice, once for each time-reversal component of the tensor. The \lstinline|symmetrize.symmetr()| function is very general and could be easily used for other purposes. One just has to provide a list of symmetries and a function that transforms the tensor by the symmetries. The tensor has to be either \lstinline|tensors.Tensor| or \lstinline|tensors.NumTensor|, however, the symmetries and the transformation function are completely general. Note also that the function assumes that the tensor is a linear combination of the independent variables and will not work otherwise.

The code supports a symbolic mode, in which the symmetrization is done fully symbolically using the Sympy \lstinline|rref()| function that transforms the matrix to the reduced row echelon form. This works well in simple cases; however, we find that in general, this approach is numerically unstable and can lead to wrong results. This approach is not fixed by using numerical transformation to the reduced row echelon form since Gaussian elimination itself is not numerically stable. Instead, by default, we do the symmetrization entirely numerically and utilize the SVD decomposition followed by a transformation to the reduced row echelon form using Gaussian elimination for the linearly independent rows. This approach works very well, and potential numerical issues in the Gaussian elimination are easily spotted since we know that the rows must be linearly independent. The disadvantage of the numerical approach is that we find the relation between the tensor components only approximately. We round the independent variable coefficients in the end, which in most cases recovers the exact relations. The numerical accuracy is controlled using the parameter \lstinline|--num-prec|, which is by default 1e-3.

The program is very fast for simple cases; however, the computational complexity grows rapidly with the rank of the tensor being symmetrized. An example of the scaling is given in Fig. \ref{fig:structures}(c) for the case of conductivity expansion in Fe. Here, a command \lstinline|symmetr res j E -f Fe.in --exp{n}| was used, where $n$ is the expansion order. In this case, the rank of the tensor being symmetrized is $n+2$. The scaling is roughly exponential, which makes it impossible to symmetrize tensors of very large rank; however, such tensors are not commonly utilized. For tensors below rank 6, the program finishes within a few minutes. It is possible to speed up the code using the flag \lstinline|--generators|. When this flag is turned on, the program first determines the generators of the symmetry group. The generators are sufficient to determine the full symmetry of the tensor because when the tensor is symmetrical under two generators $G_1$ and $G_2$, it must also be symmetric under their product $G_1G_2$. The generators are determined by gradually adding symmetries to a generator set and eliminating all symmetries that can be obtained as a product of the generators. As can be seen in Fig. \ref{fig:structures}(c), the \lstinline|--generators| flag provides a significant speedup for higher ranks, although the scaling remains the same.

\section{Examples}

This code has been used in a number of publications. Below, we give two examples of how the code was applied. The corresponding input files are included as Supplementary Material.

\subsection{Spin-orbit torque in antiferromagnets}

The spin-orbit torque is a torque acting on magnetic moments induced by electric current (or more precisely, the electric field that creates the current), which originates from the spin-orbit coupling and broken inversion symmetry \cite{Manchon2019}. In antiferromagnets, the torque has to be studied on each magnetic sublattice, which requires evaluating the symmetry corresponding to specific atoms \cite{Zelezny2017,Manchon2019}. The torque is typically described by an effective magnetic field such that $\mathbf{T}^a = \mathbf{M}^a \times \mathbf{B}^a$, where $\mathbf{T}^a$ is the torque acting on a magnetic moment $\mathbf{M}^a$ and $\mathbf{B}^a$ is the effective magnetic field. The effective magnetic field can be described by a response tensor $\chi_a$ such that $B^a_{i} = \chi_{ij}^a E_j$.

\begin{figure}
    \centering
    \includegraphics[width=0.8\linewidth]{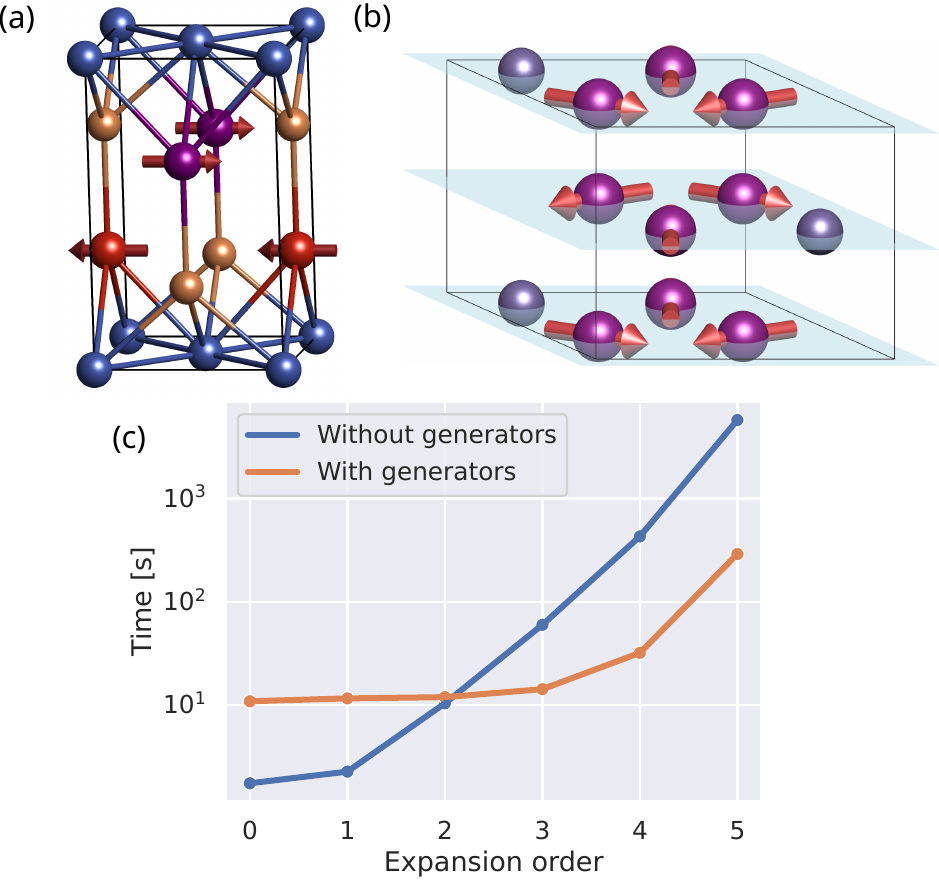}
    \caption{(a) The crystal structure of CuMnAs. Blue atoms are Cu, red and purple are Mn, and brown are As. (b) Crystal structure of Mn$_3$Sn. Purple atoms are Mn, grey denotes Sn. (c) Scaling of the running time for the expansion of the order-parameter conductivity of Fe. The blue and orange lines correspond to the case without and with the \lstinline|--generators| flag, respectively.}
    \label{fig:structures}
\end{figure}

We have studied this in the past, for example, in the antiferromagnet CuMnAs (shown in Fig. \ref{fig:structures}(a)) \cite{Wadley2017}. We use the following command to determine the symmetry of the effective field on the two magnetic sublattices:

\begin{verbatim}
    symmetr res B E -f CuMnAs.in -p 5 -p2 6
\end{verbatim}
which gives the following tensors:
\begin{align}
    \chi^{A}_\text{even} &= \left(\begin{matrix}0 & x_{01} & 0\\x_{10} & 0 & 0\\0 & 0 & 0\end{matrix}\right), \qquad
    \chi^{B}_\text{even} = \left(\begin{matrix}0 & - x_{01} & 0\\- x_{10} & 0 & 0\\0 & 0 & 0\end{matrix}\right) \\
    \chi^{A}_\text{odd} &=\left(\begin{matrix}0 & 0 & x_{02}\\0 & 0 & 0\\x_{20} & 0 & 0\end{matrix}\right), \qquad
    \chi^{B}_\text{odd} = \left(\begin{matrix}0 & 0 & x_{02}\\0 & 0 & 0\\x_{20} & 0 & 0\end{matrix}\right) 
\end{align}
These show that the $\cal{T}$-even effective field is staggered, i.e., opposite on the two sublattices. Such a field can lead to efficient manipulation of the antiferromagnetic order, which has been observed experimentally \cite{Wadley2017}. In contrast, the $\cal{T}$-odd field is uniform and does not efficiently couple to the antiferromagnetic order.
\subsection{Spin-polarized current in non-collinear antiferromagnets}

In ferromagnets, electric current has a spin-polarization. When this spin-polarized current is injected into a different ferromagnet with misaligned magnetization, the spin is absorbed, thus creating a torque, known as the spin-transfer torque. In simple high symmetry antiferromagnets, spin-polarization of the current is prohibited by symmetry; however, in antiferromagnets with lower symmetry, spin-polarized current can occur. We have shown this, for example, in non-collinear antiferromagnet Mn$_3$Sn \cite{Zelezny2017b}. The origin of this spin-polarized current is, like in the ferromagnetic case, non-relativistic.

To get the symmetry of the spin-polarized current, we use the command

\begin{verbatim}
    symmetr res s.v E -f Mn3Sn.in --noso
\end{verbatim}

This gives us the following tensor for the $\cal{T}$-odd part:

\begin{align}
    \sigma^x &= \left(\begin{matrix}0 & - x_{100} & 0\\- x_{100} & 0 & 0\\0 & 0 & 0\end{matrix}\right)\notag \\
    \sigma^y &= \left(\begin{matrix}x_{100} & 0 & 0\\0 & - x_{100} & 0\\0 & 0 & 0\end{matrix}\right) \\
    \sigma^z &= \left(\begin{matrix}0 & 0 & 0\\0 & 0 & 0\\0 & 0 & 0\end{matrix}\right)\notag
\end{align}

\section{Conclusions and outlook}

Symmetr can be used to solve the symmetry of a large array of problems in condensed matter physics. It can describe most response tensors as well as the magnetic interaction, both in the conventional case and in the non-relativistic limit. The main focus is on magnetic systems; however, non-magnetic systems can be considered as well. Other types of tensors than those currently implemented can be easily added; one only needs to specify the transformation rules of the tensor. The Python API allows integrating the program into various Python workflows. The program is relatively mature; however, other features could be implemented in the future. Currently, the expansions are implemented only for collinear systems. Implementing this feature for non-collinear systems would be useful; however, that would require significant changes to how this feature is implemented, since non-collinear systems have more than one magnetic parameter. For large tensors, the code becomes inefficient due to the exponential scaling. This could be improved by using, for example, Cython for the numerically intensive parts and potentially also by parallelizing matrix manipulation parts of the code, such as the SVD decomposition using ScaLapack or a similar library.

\section{Acknowledgement}

We acknowledge support from the Dioscuri Program LV23025 funded by MPG and MEYS, MEYS grant No. CZ.02.01.01/00/22\_008/0004594, GACR grant 25-18244S. This work was supported by the Ministry of Education, Youth and Sports of the Czech Republic through the e-INFRA CZ (ID:90254).





\bibliographystyle{elsarticle-num}
\bibliography{refs}







\end{document}